\documentclass{aa}
\usepackage{graphicx}
\usepackage{txfonts}

\usepackage{hyperref}

\begin{document}

\title{Phase-resolved optical spectroscopy of the rapidly varying white dwarf ZTF\,1851$+$1714}\titlerunning{Phase-resolved optical spectroscopy of the rapidly varying white dwarf ZTF\,1851$+$1714}

\author{
C.~C.~Pedersen\inst{1},
M.~R.~M.~Knudsen\inst{2,3},
K.~Valeckas\inst{2,3,6},
L.~Izzo\inst{4,5},
T.~M.~Tauris\inst{1},
J.~P.~U.~Fynbo\inst{2,3}
}

\institute{
Department of Materials and Production, Aalborg University, Fibigerstræde 16, 9220 Aalborg, Denmark
\and
Niels Bohr Institute, University of Copenhagen, Jagtvej 128, 2200 Copenhagen N, Denmark
\and
Cosmic DAWN Center
\and
INAF, Osservatorio Astronomico di Capodimonte, Salita Moiariello16, I-80131Napoli, Italy
\and 
DARK, Niels Bohr Institute, University of Copenhagen, Jagtvej 128, 2200 Copenhagen N, Denmark
\and
Nordic Optical Telescope, Rambla José Ana Fernández Pérez 7, local 5, E-38711 San Antonio, Breña Baja, Santa Cruz de Tenerife, Spain
}
\authorrunning{Pedersen et al.}

\date{Received 2025; accepted, 2025}

\abstract{
We report on phase-resolved optical spectroscopy and photometry in the R and B~bands of the white dwarf candidate
ZTF\,185139.81$+$171430.3. The source has been reported to be variable with a large amplitude of close to 1 magnitude, in the R~band, and a short period of 12.37~min.
We confirm this period and interpret it as the spin period of the white dwarf. The optical spectrum shows 
emission lines from hydrogen and helium superposed on a featureless continuum. The 
continuum changes shape throughout a cycle, such that it is redder when the source is bright. There is tentative evidence for Doppler shifts in the emission lines 
during the spin cycle with an amplitude of a few tens of km s$^{-1}$. 
Notably, the H$\alpha$ and H$\beta$ lines exhibit different radial velocity amplitudes, suggesting that they come from different emission regions. 
We also identify a candidate orbital period of {1.00~hr}, based on potential orbital sidebands.
These features -- Doppler shifts modulated at the spin frequency, brightness variations, and continuum shape changes -- are consistent with the accretion curtain model, in which material is funneled from a truncated inner disc along magnetic field lines onto the magnetic poles of the white dwarf.
}    


\maketitle

\section{Introduction}     
\label{sec:introduction}

Cataclysmic variables (CVs) are close binary systems comprising a white dwarf (WD) accreting material from a companion star, typically a late-type main-sequence star \citep{Robinson1976,Angel1978}. Among CVs, magnetic systems are categorised based on the strength of the WD’s magnetic field, influencing the accretion dynamics and resulting observational characteristics. Two primary sub-classes are polars and intermediate polars, distinguished by their magnetic field intensities and accretion mechanisms \citep[e.g.,][and references therein]{2008ApJ...684..558P}.

In polars, also known as AM Herculis systems, the WD possesses a strong magnetic field (typically of 10–100 MG) that prevents the formation of an accretion disc. Instead, material from the donor star is channeled along magnetic field lines directly onto the magnetic poles of the WD. 
Furthermore, the spin and orbital periods are synchronised due to the magnetic field of the WD extending to, and coupling with that of, the secondary star. Typical spin and orbital periods are $\sim$1.3--8~hr \citep[some polars have been found to be slightly asynchronous with $P_{spin}/P_{orb}\sim 0.8028\text{--}1.0028$, see][]{polar_cata}. 
The accretion process leads to high-velocity infall, producing distinctive optical spectra characterised by strong emission lines, particularly of hydrogen and helium, and significant polarisation due to cyclotron radiation. Radial velocity studies in polars reveal complex line profiles influenced by the magnetic accretion geometry and the orbital motion of the binary components \citep{Oliveira_1, Oliveira_2}. See \cite{Cropper} for an overview of polars.

Intermediate polars (IPs), or DQ Herculis stars, feature WDs with moderately strong magnetic fields (1–10 MG) and unlike for the polars, the spin and orbital period are asynchronous. Their orbital periods lie in the range $\sim$1.35--15.42~hr\footnote{https://asd.gsfc.nasa.gov/Koji.Mukai/iphome/iphome.html}, and the spin periods are often $P_{spin}\lesssim0.1P_{orb}$ \citep{Mukai_2017}. In these systems, an accretion disc may form but is truncated at the inner regions by the magnetic field. The material is then funneled along magnetic field lines onto the WD’s magnetic poles. Optical spectra of IPs exhibit prominent emission lines, including \ion{He}{ii} 4686~{\AA}, indicative of high-energy processes near the WD’s surface. Radial velocity measurements in IPs often show periodic variations that correspond to both the orbital period and the spin period of the WD, reflecting the asynchronous rotation typical of these systems \citep{TheDQHer}.

Recent discoveries have introduced the concept of ‘WD pulsars’ exemplified by systems like {AR Scorpii and J191213.72$-$441045.1} \citep{Marsh2016, 2023NatAs}. In these binaries, the WD exhibits rapid rotation and a strong magnetic field, leading to radio pulsar-like emissions powered by the WD’s spin-down energy rather than accretion.
However, interaction with the companion seems to be necessary to produce the pulse mechanism, unlike for regular pulsars.
This {suggests} that a WD-pulsar is a result of CV evolution, as is {proposed} by \citet{2021NatAs}.
To emphasise this nature, we mention the specifics of AR Sco, which is a binary system composed of a WD and an M5 dwarf with 
an orbital period of 3.56~hr. There is also a 1.95~min variation in the flux from
AR Sco, which is interpreted as the spin period of the WD. The strong magnetic field of the
WD is believed to generate synchrotron emission via charged particles coming from 
the companion M-dwarf.

\citet{Kato2021} discuss the nature of the variable source ZTF\,185139.81$+$171430.3 (ZTF\,1851$+$1714), which 
was first detected by the Zwicky Transient Factory \citep{Ofek2020}. The source 
has a short period of only 12.37~min and a large amplitude in its photometric light curve 
of almost 1 magnitude as
reported by \citet{Kato2021}. The short period and large amplitude make the
source rather unique and raise a question about its nature. 
They argue that the source most likely is a WD pulsar similar to AR Sco \citep{Marsh2016}, where
the 12.37~min variability for ZTF\,1851$+$1714, in a similar manner, reflects the spin
period of a WD in a binary. ZTF\,1851$+$1714 has also been detected 
in the X-rays, where it has been found to have a luminosity 400 times larger than 
that of AR Sco \citep{Klingler2021}.

To investigate the nature of the system, we decided to perform phase-resolved optical 
spectroscopy and photometry in two bands of ZTF\,1851$+$1714 as a project during a summer 
school at the Nordic 
Optical Telescope in August 2024. To our knowledge, no spectroscopic observations 
of ZTF\,1851$+$1714 have been presented in the literature prior to our study.

\section{Observations and data reduction}    \label{sec:data}

The main objective of our observing campaign was to secure phase-resolved spectroscopy
of the system. Given the size of the telescope (main mirror 2.56-m) and the magnitude of 
the source (ranging between 18th and 19th magnitude), we were 
restricted to using low-resolution
long-slit spectroscopy. We observed with the instrument ALFOSC and grism 19, which with 
a 1\farcs0 slit provides an average resolution of 970 (or 300~km s$^{-1}$ full-width-at-half-maximum, FWHM) 
over the spectral range
from 4400~{\AA} to 6770~{\AA}\footnote[2]{\label{spec_resolution}{https://www.not.iac.es/instruments/alfosc/grisms/}}. To reduce read-out time we decided to bin the detector by
a factor of 2 along the dispersion axis and to only read out a 500 pixel wide region along 
the slit. Before observing, we precisely timed the duration of the readout of the CCD for
this setup so that we
could calculate the exposure time that would ensure that after five spectra, we would be
back at the same position in the phase, assuming a period of 12.37~min \citep{Kato2021}. The measured read-out time was $8.38 \pm 0.01$~s. 
For five exposures, the exposure time was calculated by dividing the estimated period by the number of exposures and subtracting the measured read-out time, resulting in an exposure time of 140~s.

We also observed the target with two other grisms (18 and 20) to be able to characterise the optical
spectrum from about 3700~{\AA} into the near-infrared at 9700~{\AA}. The average spectral resolution
of these grisms are 1000 and 770, respectively, again using a 1\farcs0 wide slit\textsuperscript{\ref{spec_resolution}}.

{The slit was in all observations aligned with a nearby non-variable star, in order to have a comparison source for removing any trends from the individual spectra, related to changing airmass, seeing and similar effects, during the cycles. 
This comparison star is used in the analysis presented in Section~\ref{sec:variability}, where the target spectra were normalised by the mean flux of the comparison star in the 5000--5500~{\AA} region and subsequently scaled to match the flux level of the comparison star in the first group within the same region.} 

The observations were carried out at low airmass (see Table~\ref{tab:log}) and seeing was approximately 0\farcs65 but varied in the interval 0\farcs60--0\farcs74.
The autoguider of the telescope keeps the source in a fixed position within the
slit to a precision that is well below 0\farcs1, and therefore we do not expect spurious radial velocity signals from the source moving within the slit.
The log of observations can be inspected in Table~\ref{tab:log}, and the precise start times
for the 40 spectra obtained with grism 19 can be found in Table~\ref{tab:grouped_times}.

We secured photometric light curves on two epochs in the Bessel B and R filters in order to examine the colour behaviour of the photometric variability. 
The observations were reduced using aperture photometry, and brightness variations were measured via differential photometry using the nearby comparison star. The exposure times and cadence of these observations are listed in Table \ref{tab:log}.

\begin{table}[!t]
\centering
\begin{minipage}{0.5\textwidth}
\centering
\caption{Log of ALFOSC observations in August 2024.}
\begin{tabular}{lllcc}
\noalign{\smallskip} \hline \hline \noalign{\smallskip}
Observation & Date  & Exposure time & Airmass\\
       &       &   (s)      \\
\hline
grism 19 & 08/08/2024 & 40$\times$140 & 1.061--1.020 \\
grism 18  & 24/08/2024 & 2$\times$600 & 1.028--1.023 \\
grism 20  & 10/08/2024 & 300  & 1.064 \\
Bessel B & 23/8/2024  & 70$\times$50 & 1.020--1.037 \\
Bessel R & 25/8/2024  & 70$\times$50 & 1.020--1.051 \\
\hline
\noalign{\smallskip} \hline \noalign{\smallskip}
\end{tabular}
\centering
\label{tab:log}
\tablefoot{The exposure times indicated are the integration times only.}
\end{minipage}
\end{table}

\begin{table}[h!]
\centering
\begin{minipage}{0.5\textwidth}
\caption{Start Universal Time (UT) for each of the 40 grism 19 exposures ordered by group, i.e. by phase.} 
\begin{tabular}{lllcc}
\noalign{\smallskip} \hline \hline \noalign{\smallskip}
\# & Start times for exposures  \\ \hline
1 & 21:43:58.206, 21:56:20.014, 22:08:41.832, 22:21:03.623 \\ 
  & 22:33:25.552, 22:45:47.436, 22:58:09.234, 23:10:31.057 \\ \hline
2 & 21:46:26.552, 21:58:48.352, 22:11:10.180, 22:23:32.019 \\
  & 22:35:53.962, 22:48:15.775, 23:00:37.606, 23:12:59.456 \\ \hline
3 & 21:48:54.948, 22:01:16.739, 22:13:38.549, 22:26:00.409 \\
  & 22:38:22.312, 22:50:44.152, 23:03:05.966, 23:15:27.792 \\ \hline
4 & 21:51:23.292, 22:03:45.092, 22:16:06.885, 22:28:28.756 \\
  & 22:40:50.683, 22:53:12.530, 23:05:34.340, 23:17:56.169 \\ \hline
5 & 21:53:51.670, 22:06:13.433, 22:18:35.260, 22:30:57.154 \\
  & 22:43:19.062, 22:55:40.893, 23:08:02.698, 23:20:24.511\\
\hline
\noalign{\smallskip} \hline \noalign{\smallskip}
\end{tabular}
\label{tab:grouped_times}
\tablefoot{All times are for August 8 2024. The difference between two subsequent observations is the exposure time (140~s), the readout time (8.38~s) and the (negligible) computational overhead for the period that the observation system takes to end an observation and initiate the next.}
\end{minipage}
\end{table}

The data were reduced using standard methods for wavelength calibration, extraction, 
and flux calibration in longslit spectroscopy as 
implemented in a Python code \citep{Valeckas2025}. Flux calibration was 
done using observations of spectrophotometric standard stars observed on the 
same nights as the science data. 

\section{Results}    \label{sec:results}

\subsection{The full spectrum}
\begin{figure*}[th]
\centering
\includegraphics[scale=0.54]{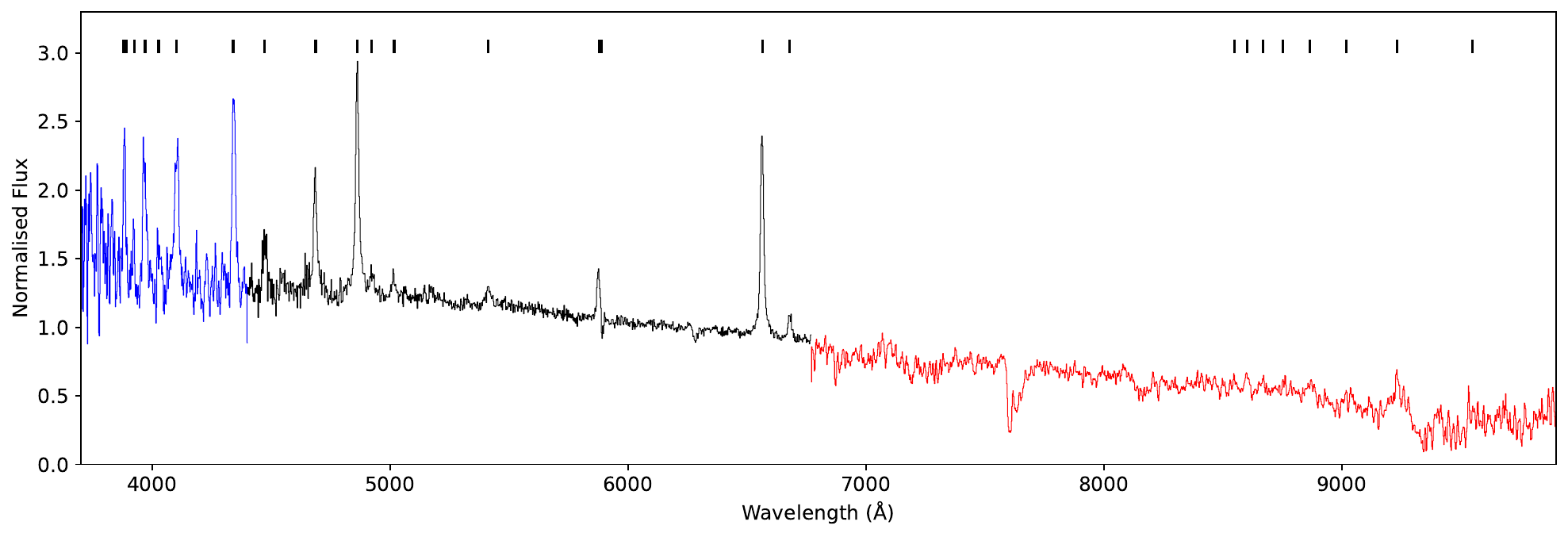}
\caption{
Stacked spectrum using all exposures in all grisms. The blue section represents
the grism 18 observation, the black spectrum the weighted mean of the 40 grism 19 spectra,
and the red section represents the grism 20 spectrum. The positions of emission lines
from \ion{H}{i}, \ion{He}{i}, and \ion{He}{ii} are indicated with vertical lines at the top. Some of the 
hydrogen lines in the grism 20 spectrum are only tentatively detected. We
also mark the position of NaD absorption, which is presumably interstellar in nature.
The spectrum has been normalised to 1 at 6000~{\AA}.}
\label{fig:fullspectrum}
\end{figure*}

In Fig.~\ref{fig:fullspectrum}, we show the full optical spectrum covering the region of spectra from near-ultraviolet to near-infrared using spectra from
all three grisms.
As the object is likely far out (parallax of $0.325 \pm 0.119$~mas \citet{Kato2021}), we have corrected the spectrum for a full line-of-sight foreground extinction in the direction of 
ZTF\,1851$+$1714 of $A_\mathrm{V} = 0.996$ mag as determined from the extinction map of 
\citet{Schlafly2011}. The spectra taken with grism 19 were RV-corrected prior to being averaged. 
The spectrum is characterised by a large number of emission lines from \ion{H}{i}, \ion{He}{i}, and
\ion{He}{ii} superposed on a fairly featureless continuum.
The emission lines are resolved in velocity. 
Also the H$\alpha$ and H$\beta$ lines are prominent, where for H$\alpha$ we measure a Gaussian width
of $7.3\pm0.3$~{\AA} corresponding to 750~km~s$^{-1}$ FWHM
corrected for the spectral resolution.
There are absorption features
at 5890~{\AA} and 6280~{\AA} (in addition to absorption at the telluric A and B~bands). 
{The 5890~{\AA} is presumably caused by interstellar Na, as no significant periodic variability is measured in its radial velocity; however, this interpretation remains tentative given the low spectral resolution and signal-to-noise ratio.
The 6280~{\AA} line is caused by telluric absorption due to oxygen.}

The presence of prominent hydrogen lines makes it unlikely that the 12.37~min variability reflects the orbital period. 
As we understand, CVs with hydrogen-rich donors have an observed orbital period minimum of approximately $78\text{--}82$~min \citep{Gansicke2009}. 
This minimum may be pushed to $\sim$51~min for extremely low-metallicity donors \citep{Stehle1997}, and potentially as low as $\sim$37~min if the companion is a degenerate object such as a brown dwarf or gas giant \citep{Rappaport_2021}. 
Periods significantly below these thresholds require highly compact donors and are typically associated with helium dominated systems such as AM~CVns, which typically show no hydrogen in their spectra.
While trace amounts of hydrogen have been detected in the spectra of some AM~CVn-type systems (e.g. HM Cnc; see \citealt{HM_cnc_2002,HM_cnc_2022}), the amount of hydrogen is estimated to be very low and the optical spectra remain helium dominated.

In the case of ZTF\,1851$+$1714, the 12.37~min variability is well below these period limits, yet the optical spectrum is hydrogen-rich. 
This strongly suggests that the periodicity does not represent the orbital period of the system. Instead, we interpret it as the spin period of the WD.

\subsection{Phase-resolved spectroscopy}
\begin{figure}[th]
\centering
\includegraphics[width=1.0\columnwidth]{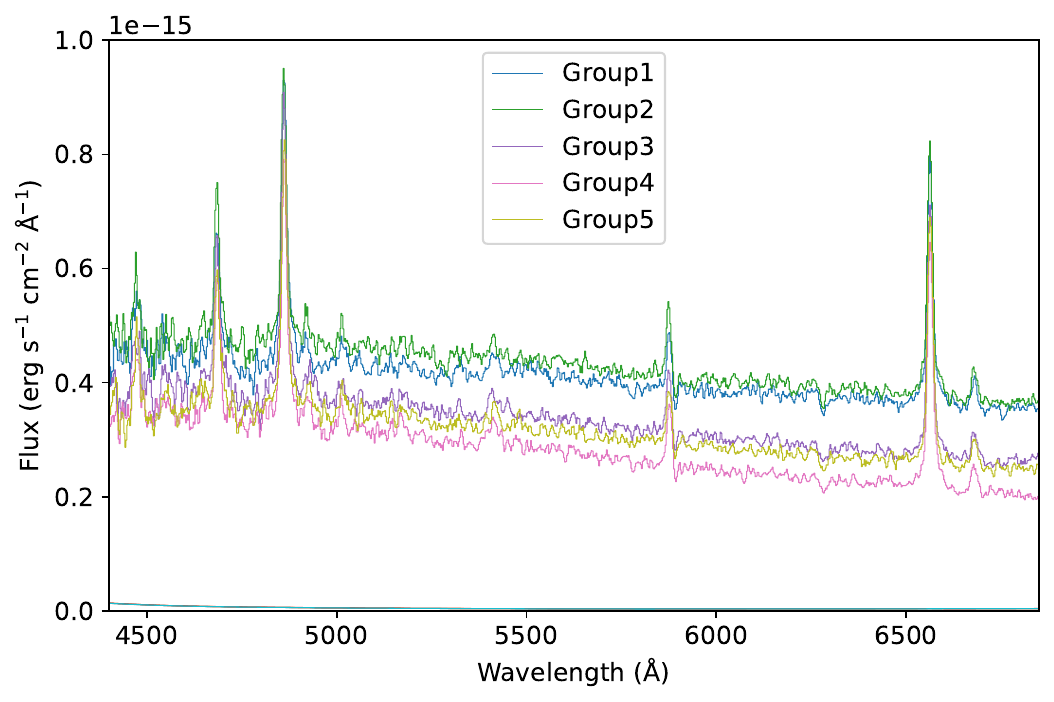}
\caption{
Phase-resolved spectroscopy covering the spectral region from 4400~{\AA} to 6770~{\AA} 
using grism 19. We show the stacked spectra for each of the given groups covering the 
full phase of the 12.37~min variability period.}
\label{fig:spectra_phased}
\end{figure}

The phase-resolved grism 19 spectra can be seen in Fig.~\ref{fig:spectra_phased}. 
Here, we have stacked the spectra in five groups ordered by phase. Note that start time is arbitrary as we did not synchronise the start of the observations with the ephemeris
of the variable as these are not known well enough to allow this. 
The spectra show a strong chromatic evolution during the phase: the continuum is significantly redder when the 
source is bright compared to the spectrum closer to minimum flux. 

\begin{figure}[th]
\centering
\includegraphics[width=1.0\columnwidth]{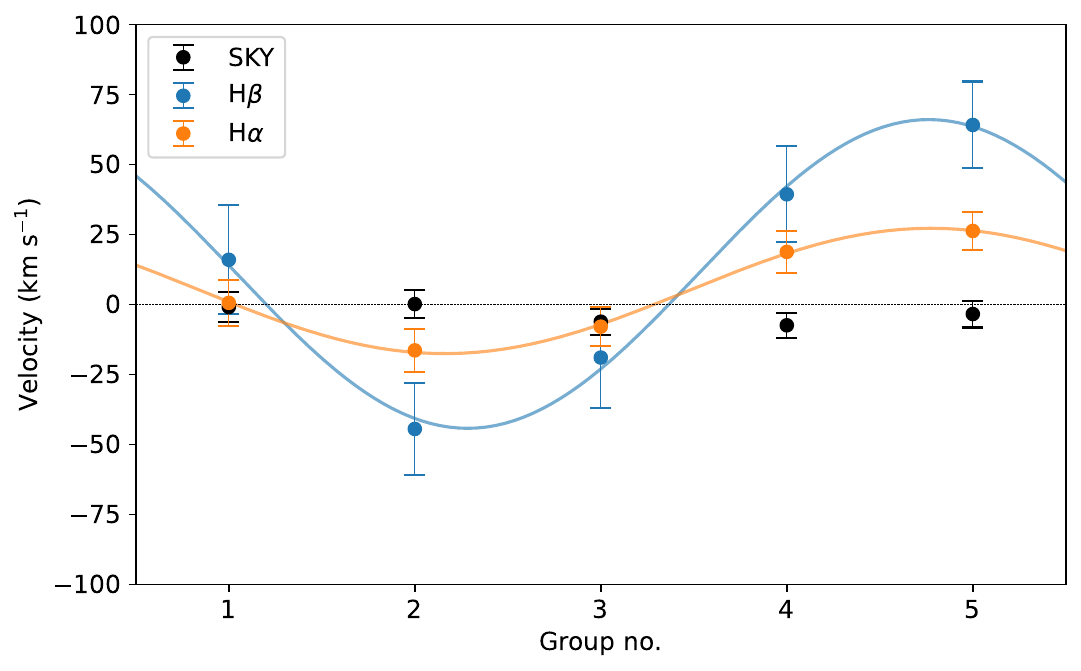}
\caption{
Velocities derived from Doppler shifts measured from the H$\alpha$ and H$\beta$ lines covering the 
full phase of the 12.37~min variability period (see Fig.~\ref{fig:spectra_phased}).
Sine curves have been fitted to the H$\alpha$ and H$\beta$ data points to highlight their periodic nature. 
For comparison, we also show velocities measured from
sky-lines in the same spectra. }
\label{fig:velocities}
\end{figure}

We also measure the Doppler shifts of the emission lines. Only the H$\alpha$ and H$\beta$ lines have sufficient signal-to-noise ratio to allow a significant measurement, but for these two lines, we find fairly consistent results for all the spectra (see Fig.~\ref{fig:velocities}).
The radial velocity of both of these lines vary sinusoidally with the observed period, meaning that they are related to the spin of the WD. 
This could indicate that they originate from where the material of the accretion disc is able to couple to the magnetic field of the WD and hence rotate with it.
A difference in radial velocity amplitude is also apparent, which means that the two lines presumably originate from different places.

\subsection{Equivalent width of the H$\beta$ line}
An empirical criterion for classifying magnetic CVs has also been suggested by \citet{Silber1992PhDT}.
According to this criterion, the equivalent width (EW) of the H$\beta$ line should be greater than 20 {\AA} and the ratio between the amplitude of \ion{He}{ii} 4686 {\AA} / H$\beta > 0.4$.
Calculating these for each of the five spectra in Fig.~\ref{fig:spectra_phased}, using Specutils \citep{specutils}, yields a mean EW of $29 \pm 4$ {\AA} and a mean ratio \ion{He}{ii} 4686 {\AA} / H$\beta$ of $0.72\pm 0.04$.
These values are above the thresholds, even when considering errors, implying that ZTF 1851+1714 is a magnetic CV. 

\subsection{Variability analysis} \label{sec:variability}
In order to investigate the periodic variability in the magnitude of ZTF 1851+1714, we also looked at the 40 individual spectra and R and B~band photometry. 
{The spectra were produced in the same manner as the five spectra in Fig.~\ref{fig:spectra_phased}, but the comparison star was used to further calibrate each spectrum, and they have not been corrected for foreground extinction.}
These spectra are noisier, which is expected, and they present some of the same features, mainly the H$\alpha$ and H$\beta$ lines and periodic flux variation.

The values of the integrated fluxes can be seen in Fig.~\ref{fig:tot_period}. 
They were obtained by numerically integrating the spectra in their wavelength range, which is $\sim$4404 {\AA} to $\sim$6964 {\AA}. 
A clear sinusoidal variability is seen, corresponding to the eight periods observed. 
Another modulation, with a longer period of $\sim0.6$~hr also appears to be present, looking at the dotted blue peaks around $[0.2, 0.8, 1.4]$~hr.

To investigate the potential different modulations, a fast Fourier transform (FFT) was used to produce a power spectrum of the integrated fluxes (Fig.~\ref{fig:Power_spec}).
An oversampling factor of {20} was applied to improve frequency resolution, and the integrated flux values were normalised in terms of the maximum integrated flux, to minimise numerical errors. 
The frequency range of 3–12~hr$^{-1}$ was selected because the lower frequencies were too close to the total observation time, while the higher frequencies were close to the exposure times, making them less reliable.
{Furthermore, the power spectrum was normalised and a detection threshold corresponding to a 5\% false-alarm probability (FAP) has been included to highlight the significance of peaks.}

A clear dominant frequency is present corresponding to a period of {12.29~min}, which matches the spin period of the WD \citep{Kato2021}{, to an error within 1\%.} 
No other modulations are immediately apparent.

To further investigate the periodic behaviour, we consider the presence of orbital sidebands.
In intermediate polars, optical signals commonly exhibit orbital sidebands at $\omega - \Omega$, where $\omega$ represents the WD's spin frequency and $\Omega$ denotes the orbital frequency \citep{TheDQHer}. 
If the peak observed at 3.88~hr$^{-1}$ in the top panel of Fig.~\ref{fig:Power_spec} corresponds to this orbital sideband, it implies an orbital frequency of {$\Omega = 1.00$~hr$^{-1}$, leading to an estimated orbital period of 1.00~hr.}
Additionally, a peak at 7.76~hr$^{-1}$ is detected, which corresponds exactly to $2(\omega - \Omega)$. 
This suggests an $\sim$1~hr orbital period for the binary system, although it cannot be definitively confirmed (See Section \ref{sec:discussion} for a discussion of this orbital period compared to those of known IPs).

The aliasing structure resulting from the spectral window of the 40 grism~19 spectra is overplotted with the power spectrum in the top panel of Fig.~\ref{fig:Power_spec}. {It has been scaled by the same factor as the normalised power spectrum.} 
It can be seen that the proposed orbital sideband at 3.88~hr$^{-1}$ lies close to an alias peak introduced by the spectral window, although the two do not exactly coincide. 
The feature at 7.76~hr$^{-1}$ is located further away from the nearest alias peaks. 
This makes it difficult to determine whether this represents a true periodic signal or is an artefact from the observation window.
{However, the peaks do not reach the detection threshold, making it even more uncertain whether they are indeed from a true periodic signal.}  

\begin{figure}[th]
\centering
\includegraphics[width=0.95\columnwidth]{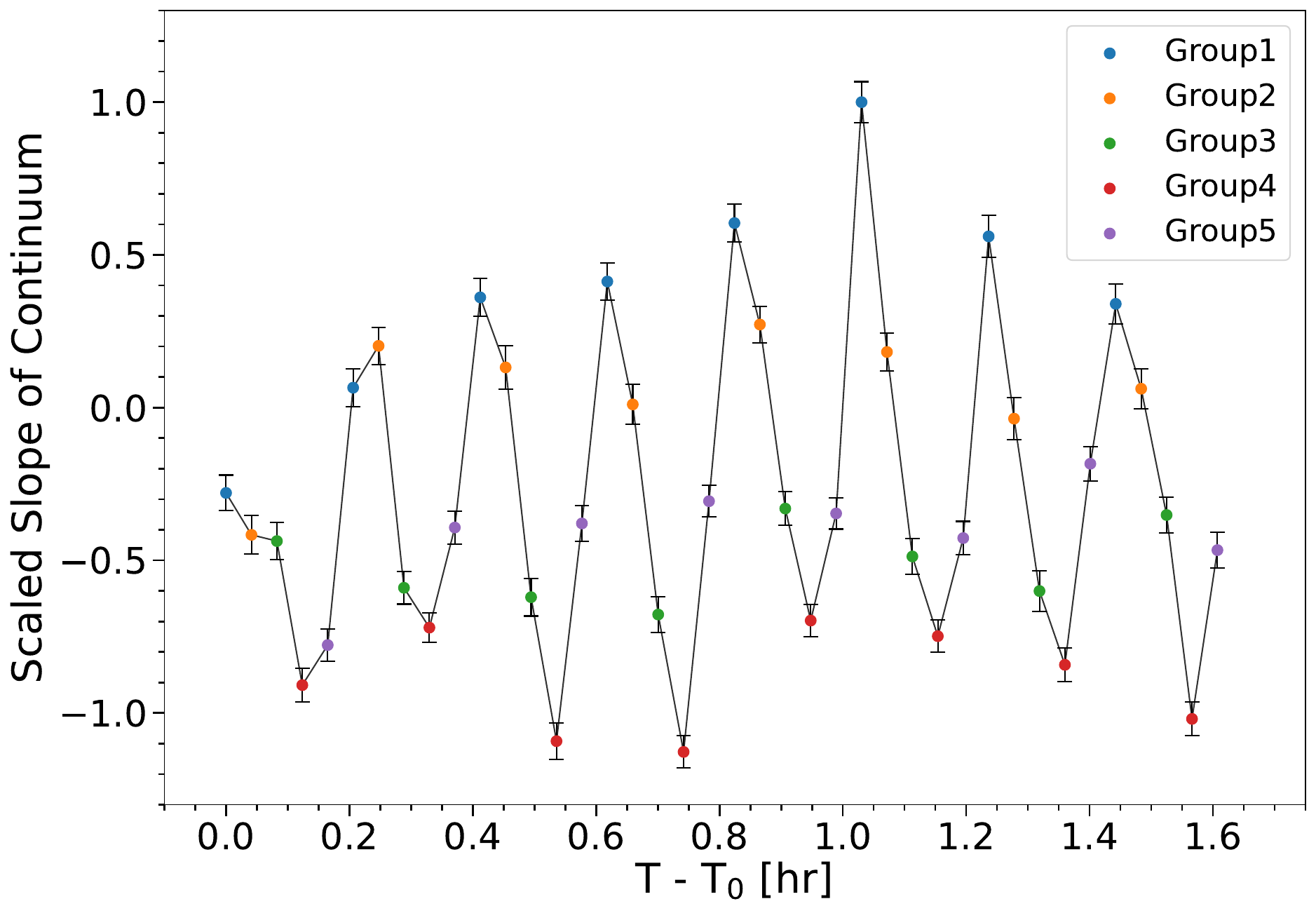}
\caption{Slopes of the continua for the different spectra, obtained by fitting a straight line to each spectrum. 
The values are scaled in terms of the continuum with the highest slope, such that this value equals one.}
\label{fig:slope_spec}
\end{figure}

To investigate the wavelength-dependent variations in the spectra, we fitted straight lines to the different continua. 
This was done to estimate the slopes, which would give us a measure of whether the spectra were more blue or red.
If the slope is negative, we have a bluer continuum, because this would mean that the flux is larger at lower wavelengths and drops off towards higher wavelengths.
When the slope is positive, we have a redder spectrum, as the flux increases with wavelength.
The resulting slope for each spectrum is shown in Fig.~\ref{fig:slope_spec}.
They have been scaled in terms of the maximum value of $1.07\times10^{-4}$~erg~s$^{-1}$~cm$^{-2}$~{\AA}$^{-2}$, to highlight the periodic variation of the slope throughout the eight observed periods.

Comparing Fig.~\ref{fig:slope_spec} with Fig.~\ref{fig:tot_period}, we find that when the integrated flux is low, the slopes tend to be negative, meaning that the spectra are bluer when there is a minimum in the integrated flux.
The opposite is seen when the integrated flux is high, the slopes tend to be positive, and hence the spectra are redder.
Additionally, a slight phase shift is present. In Fig.~\ref{fig:slope_spec} the highest integrated flux values are mainly associated with group 1, whereas in Fig.~\ref{fig:tot_period}, they are more strongly associated with group 2.

\begin{figure}[th]
\centering
\includegraphics[width=0.95\columnwidth]{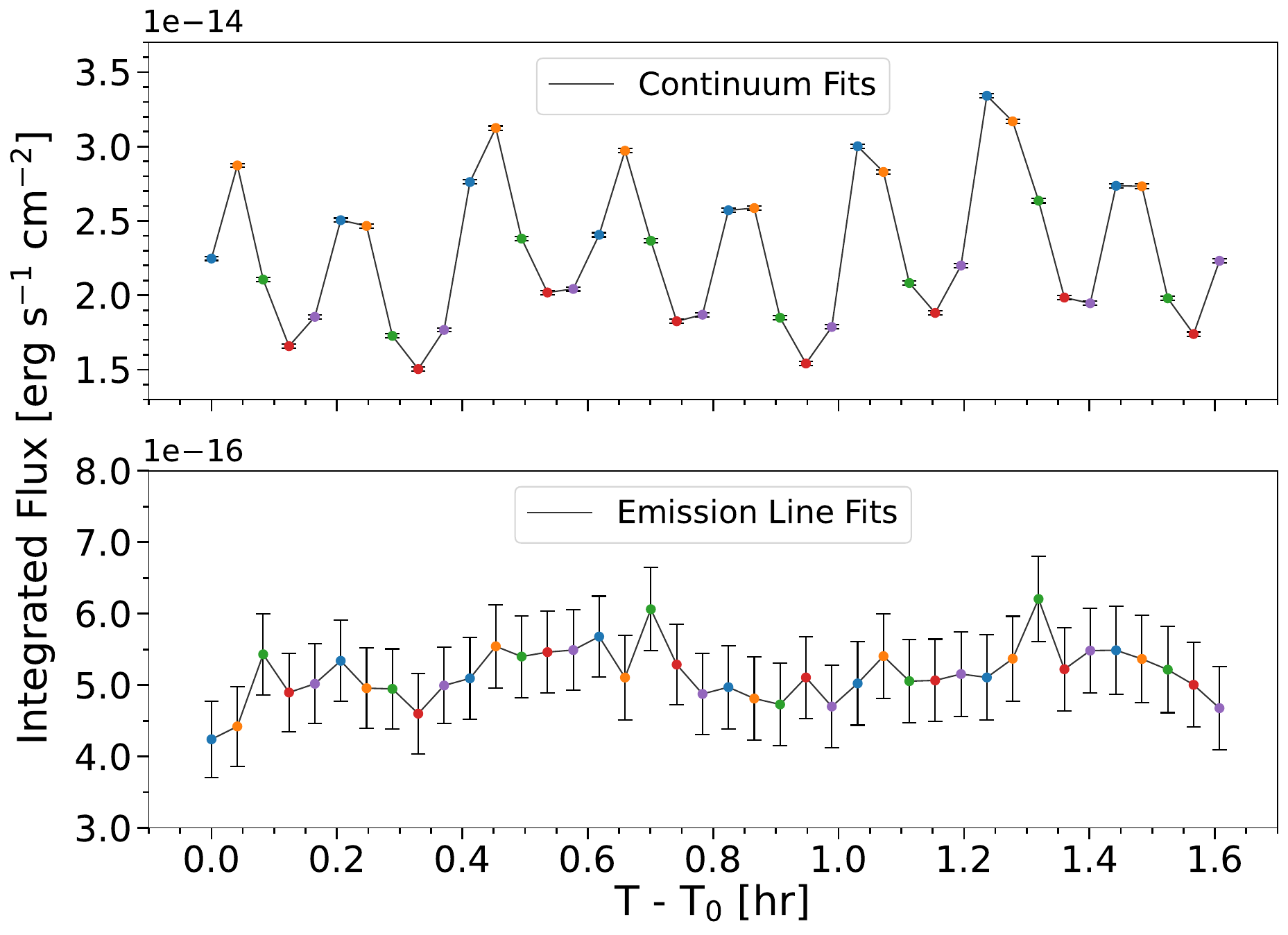}
\caption{(Top) Estimated integrated fluxes for the continuum, which has been obtained by fitting a polynomial to the parts of the spectra, without emission lines. (Bottom) Estimated integrated fluxes for the \ion{He}{ii}, H$\beta$, \ion{He}{i} and H$\alpha$ emission lines, obtained from the spectra by subtracting the continuum and fitting Gaussian functions to each of the four lines.}
\label{fig:cont_vs_emm}
\end{figure}

In order to see how the amplitude of the main four emission lines, \ion{He}{ii}, H$\beta$, \ion{He}{i} and H$\alpha$, varies compared to the continuum, we separated them from the continuum. 
This was done by fitting a polynomial to parts of the spectra without emission lines, thus obtaining a fit of the continuum. 
Subtracting this continuum from the spectra, we could fit Gaussian functions to each of the four lines.
Integrating these fits for each of the spectra, we get Fig.~\ref{fig:cont_vs_emm}.
Here, it is seen that the emission lines do not show any periodic behaviour.
However, because these lines have a much smaller area in terms of the calculated integrated flux, they are also more susceptible to noise{, which is also evident from the error bars}.
It is also seen that the periodic behaviour of the WD is dominated by the continuum since the integrated fluxes correspond to those in Fig.~\ref{fig:tot_period}.

\subsection{Photometry}
\begin{figure}[th]
\centering
\includegraphics[width=0.95\columnwidth]{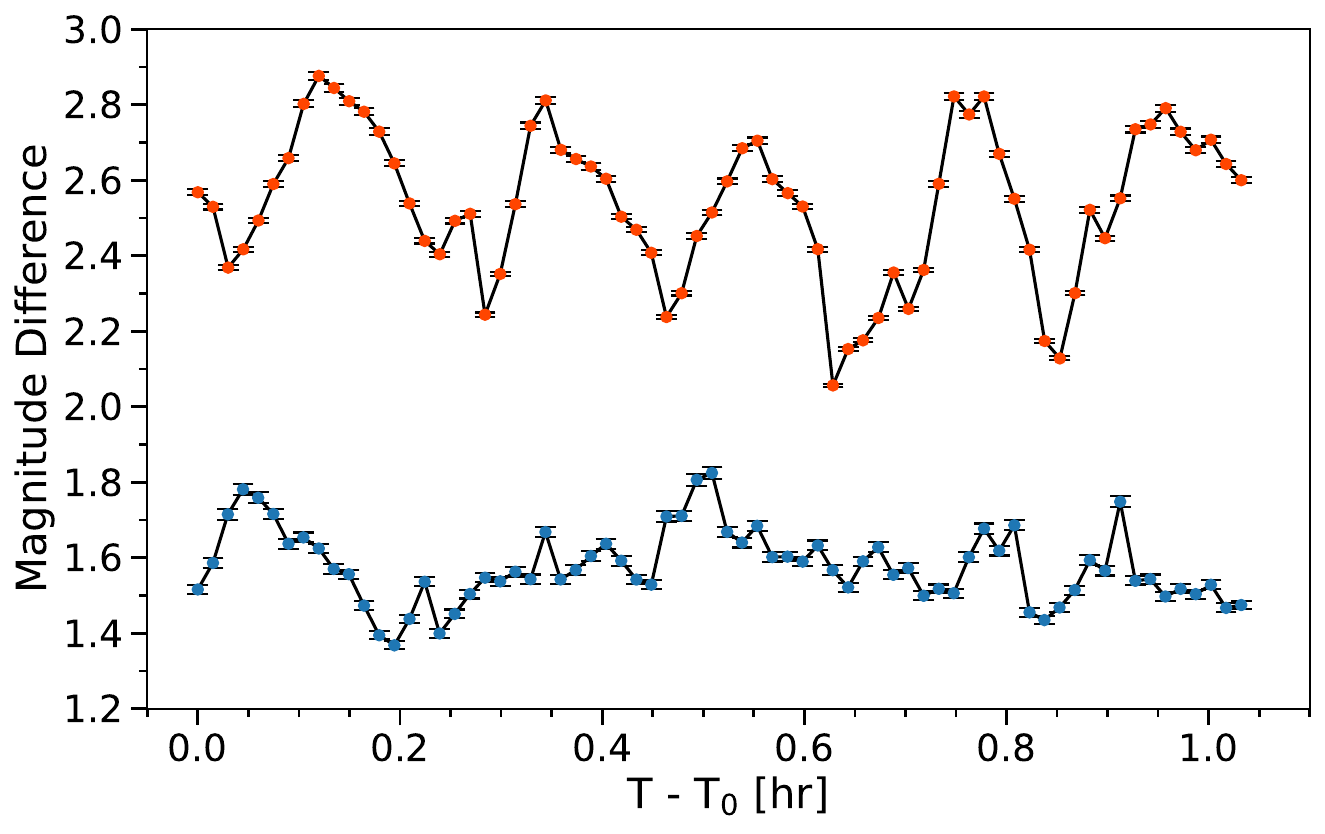}
\caption{R and B~band photometry measurements, displayed by red and blue dots, respectively. The R~band measurements were taken on August 25 2024 and the B~band measurements on August 23 2024. The magnitudes are displayed in terms of the difference between ZTF 1851+1714 and the reference star.}
\label{fig:R_B_band_phot}
\end{figure}

To build on the findings of the integrated spectra, we took photometric measurements in the R and B~bands (see Fig.~\ref{fig:R_B_band_phot}). 
This was done to compare the results and to investigate orbital sidebands by making FFT power spectra.
A total of 70 measurements were made in both bands with exposure times of 50~s, which means $\sim$1~hr of observation time, covering $\sim$ five~spin periods of the WD.

\begin{figure}[th]
\centering
\includegraphics[width=0.95\columnwidth]{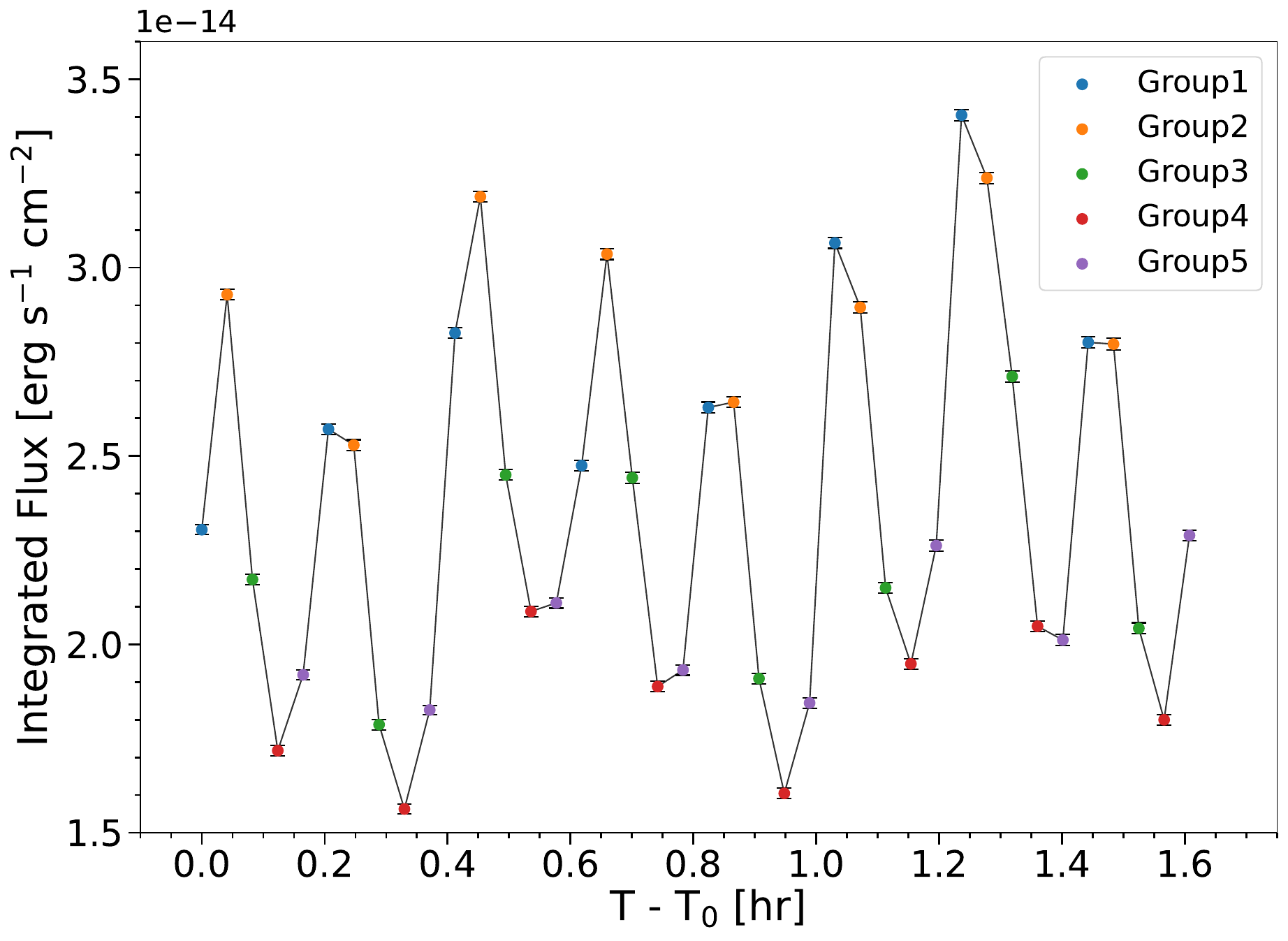}
\caption{Calculated integrated flux for each of the 40 spectra. Each spectrum has been numerically integrated in their wavelength range. The time assigned to each spectrum is the start time of the exposure. These timestamps were shifted such that the first measurement starts at 0.     
}
\label{fig:tot_period}
\end{figure}

\begin{figure}[th]
\centering
\includegraphics[width=0.95\columnwidth]{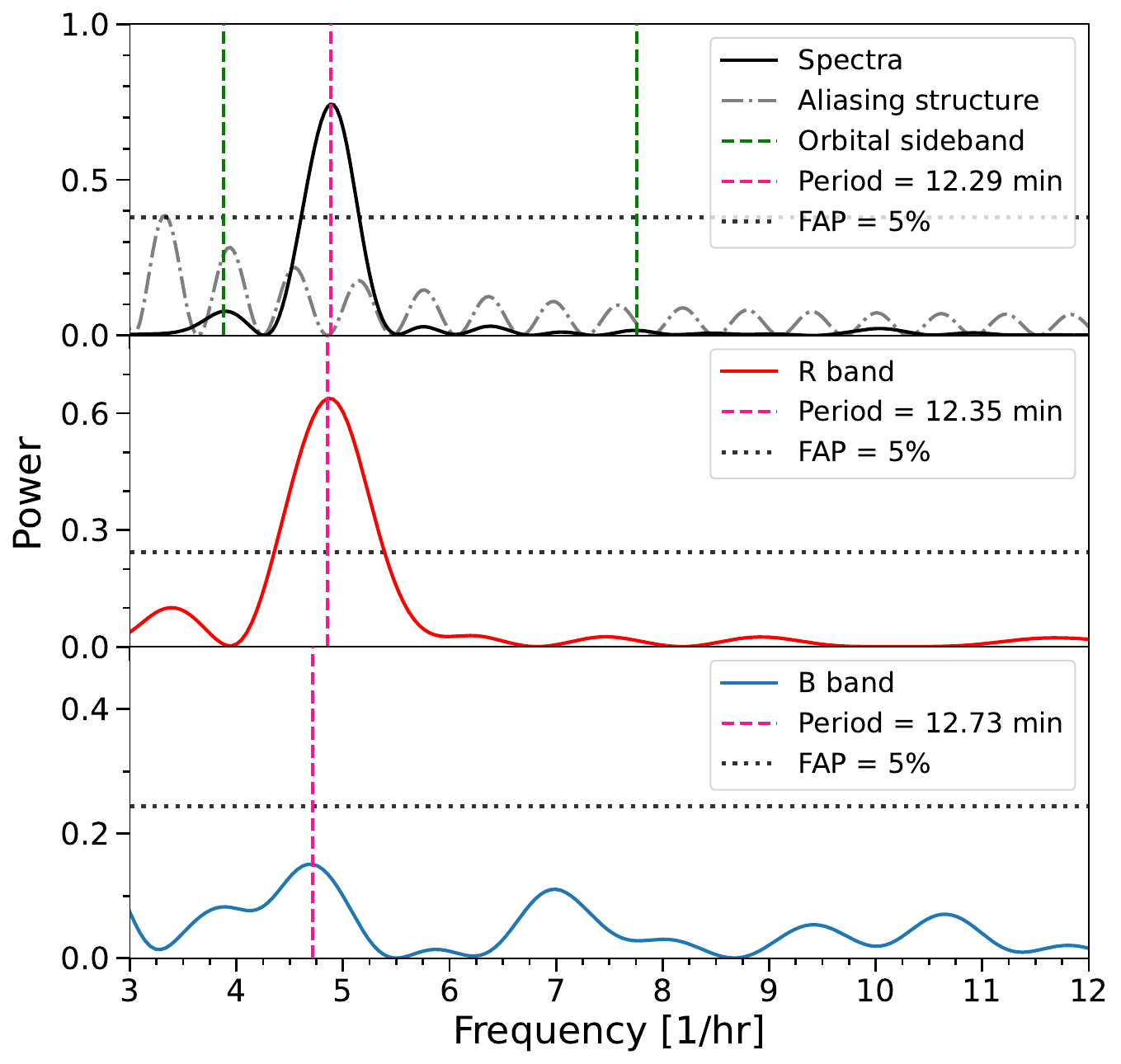}
\caption{Normalised FFT power spectra of the integrated fluxes seen in Fig.~\ref{fig:tot_period}, along with the R and B~band photometry measurements. The fluxes have been normalised in terms of the maximum flux. An oversampling factor of 20 has been used to get better frequency resolution in all three power spectra. The aliasing structure resulting from the spectral window of the observations for grism 19 is shown in the top panel, along with the potential orbital sidebands. The aliasing structure has been scaled by the same factor as the normalised power spectrum. 
A detection threshold corresponding to a 5\% FAP has been calculated for each spectrum.}
\label{fig:Power_spec}
\end{figure}

The same periodic behaviour as in Fig.~\ref{fig:tot_period} is present in the R~band, however, no other modulations are found upon visual inspection. 
Looking at the power spectrum for the R~band in Fig.~\ref{fig:Power_spec} confirms that the dominating period is the spin period of the WD.

In the B~band measurements, many different modulations are present, but the previous periodic behaviour of the WD spin is not immediately apparent. 
Examining the power spectrum, a peak was found close to the frequency corresponding to the spin period of the WD.
However, the peaks are less prominent than for the R~band measurements, making it difficult to determine anything with confidence. 
{This is also evident from the fact that none of the peaks reach the detection threshold.} 

\section{Discussion}
\label{sec:discussion}
 
The changing shape and luminosity of the continuum during the 12.37~min variability period require explanation.
If it were a hot spot on the WD, we would 
see the opposite trend: bluer spectrum correlating with a higher luminosity. If there was a cold spot we would 
expect the same trend: the luminosity would be lower when the cold spot was on the line of sight. 
One explanation could be a non-thermal source of continuum, i.e. synchrotron or cyclotron emission.
Another possibility would be occultation of a red source in the faint state, i.e. occultation of a red companion star, but the light curve does not
look like that of an eclipsing binary.

A more plausible explanation involves the accretion geometry close to the WD, where the accretion disc is truncated. 
The periodic variations in the velocities of the H$\alpha$ and H$\beta$ emission lines, derived from Doppler shifts, confirm that we are receiving light from this region, as they follow the same periodic behaviour as the WD.

In the accretion curtain scenario, the matter from the disc is funneled along the WD’s magnetic field lines, forming a curtain-like structure that channels material onto the magnetic poles. 
This model suggests that we receive more light when viewing these curtains from the side, specifically when the magnetic poles are not pointing directly towards us, because a larger emitting surface area is exposed (see Chapter 9 in \cite{2001cvs_Hellier}). 

This would imply that, when the integrated flux is at maximum, we might not see the hot spot directly, leading to a redder spectrum. 
Conversely, when the magnetic pole is directed towards us, some of the light is absorbed or scattered within the curtain, reducing the integrated flux. 
However, since the pole is facing us in this phase, we observe an increase at blue wavelengths as the hot spot is visible.

A slight phase shift is observed between the integrated flux and the radial velocities of the H$\alpha$ and H$\beta$ emission lines. Although the lines are generally blueshifted near flux maximum and redshifted near flux minimum, this correspondence is not exact. As is seen in groups 4 and 5 of Fig.~\ref{fig:spectra_phased} and Fig.~\ref{fig:velocities}, the maximum redshift occurs just after the flux minimum. This offset is consistent with the accretion curtain model; as the magnetic pole and curtain rotate and point towards us, the continuum flux drops due to increased absorption or scattering, and shortly thereafter the curtain's motion leads to a redshift, as it is moving away from us. Likewise, as the curtain rotates and points away from us, the emitting area increases, the flux reaches a maximum, and the inflowing material begins moving towards us again, producing a blue shift. A similar phase relationship is observed in the intermediate polar EX~Hydrae \citep{Mhlahlo_2007}.

Furthermore, \cite{Mhlahlo_2007} argue that the Doppler shift of the H$\beta$ line in EX~Hydrae originates from the rotation of the funnel at the outer disc edge, where a shock is created, while the Doppler shift of the narrow component of the H$\alpha$ line originates from the flow of material along the field lines in the curtain.
This could explain the amplitude difference that we see for ZTF~1851+1714, since the H$\beta$ line obtains higher radial velocities.
It is also plausible that this shock causes an increase in the received flux at the blue wavelength during minimum of the integrated flux, as it would be facing us, while at maximum the funnel overshadows it, resulting in a redder spectrum. 

In terms of the variability analysis of the integrated spectra and the photometry of the R and B~band, we find the same periodic variability as in \citet{Kato2021}, which is believed to be the WD spin period.
Additional periodic modulations at 3.88~hr$^{-1}$ and 7.76~hr$^{-1}$ were present, which could be related to an orbital sideband if the binary has an orbital period of $\sim$1.00~hr.
Although it is theoretically possible to have orbital periods down to $\sim$37~min with a brown dwarf companion \citep{Rappaport_2021}, it seems unlikely for ZTF\,1851$+$1714 to have such an orbital period, considering those of the current, classified intermediate polars, which range from 1.35--15.42~hr. 
The absence of orbital sidebands in most measurements further supports the conclusion that the modulation is not directly linked to the orbital dynamics of the binary system.  

The limited time spans of the observations also restrict our sensitivity to long-period modulations. 
To better constrain the orbital period and potential sidebands, additional observations are needed, ideally spanning multiple nights and covering several WD spin cycles to reduce aliasing effects.
A higher number of phase-resolved spectra covering the spin period of the WD would improve our ability to track spectral variations over time, allowing for a more detailed characterisation of how the observed features evolve throughout the period. 
High resolution phase-resolved spectroscopy would enable a more detailed study of emission regions and potentially allow for Doppler tomography \citep{Doppler_tomography}. Polarimetric observations could further constrain the magnetic geometry and test for spin-modulated polarisation.

Previous X-ray observations using the Swift X-Ray Telescope were reported by \citet{Klingler2021}, but these data were not examined for the presence of orbital sidebands. 
Since intermediate polars often show clear X-ray modulations at the WD spin period and potentially at sideband frequencies, re-analysis of the Swift data, or deeper follow-up with current X-ray observatories, could provide an independent check on the proposed $\sim$1.00~hr orbital period. 

\section{Conclusion}
\label{sec:conclusion}

The observational properties of ZTF~1851+1714 are best explained by a geometry consistent with an intermediate polar. 
The periodic behaviour of the continuum and emission lines and the observed colour variations all support a scenario in which accretion is controlled by a magnetic field that truncates the inner disc and channels material along field lines to the WD poles.

The phase-dependent blue and red continuum behaviour can be explained by the variable visibility of the hot spot and accretion curtain, and the behaviour of the Doppler shifted emission lines is consistent with flow patterns seen in other intermediate polars such as EX~Hydrae. 
The absence of eclipses, the short photometric period, and the spectroscopic characteristics all argue against other interpretations, such as WD pulsations or a simple binary eclipse.

Although some features, particularly the orbital period, remain uncertain, ZTF~1851+1714 shows strong similarities to known intermediate polars and is a compelling candidate for inclusion in this class.
Follow-up observations across multiple wavelengths and longer baselines will be essential to confirm this classification and further understand the structure of the system.

\begin{acknowledgements}
Based on observations made with the Nordic Optical Telescope, owned in collaboration by the University of Turku and Aarhus University, and operated jointly by Aarhus University, the University of Turku and the University of Oslo, representing Denmark, Finland and Norway, the University of Iceland and Stockholm University at the Observatorio del Roque de los Muchachos, La Palma, Spain, of the Instituto de Astrofisica de Canarias. The data presented here were obtained with ALFOSC, which is provided by the Instituto de Astrofisica de Andalucia (IAA) under a joint agreement with the University of Copenhagen and NOT. JPUF is supported by the Independent Research Fund Denmark (DFF--4090-00079) and thanks the Carlsberg Foundation for support.
The authors would like to thank the participants of the NOT Summer School 2024 who offered valuable discussions and insights that helped shape the direction of this work.
\end{acknowledgements}

\bibliographystyle{aa}
\bibliography{aa55119_25}

\end{document}